# Reconstructing a website's lost past
# Methodological issues concerning the history of www.unibo.it

Federico Nanni

The University of Bologna is considered to be the world's oldest university in terms of continuous operation. Its nine-hundred years old roots lead to the figure of Irnerius, a jurist and instructor of law of the XII Century[1].

The study of specific aspects of the history of this institution has already offered to historians a unique possibility of digging deeper in the relationship between the university, its large students' community and the city of Bologna itself[2]; moreover this kind of research has also guaranteed a better understanding of its key historical role in the Italian academic ecosystem[3]. Several sources have been used to trace its past, starting from textual documents preserved in its archives[4] to its collection of over seven hundred portraits[5].

Since the introduction of the World Wide Web, a new and different kind of primary source has been available for researchers: born digital documents, materials which exist only online and which will become increasingly useful for historians interested in studying the recent times (Webster, 2012 and 2015).

However these sources are already more difficult to preserve in their integrity compared to traditional ones (Brügger, 2008) and at the same time they are often too many to be analyzed without employing a computational approach (Graham, Milligan and Weingart, 2013).

---

[1] Brizzi Gian Paolo et al., *L'Università a Bologna*. Cassa di Risparmio in Bologna, 1988.
http://www.unibo.it/en/university/who-we-are/our-history/university-from-12th-to-20th-century

[2] Barbagli Marzio, Colombo Asher, Orzi Renzo. *Gli studenti e la città: primo rapporto sugli studenti dell'Università di Bologna,* Bologna University Press, Bologna, 2009.
Brizzi Gian Paolo. "La presenza studentesca nelle università italiane nella prima età moderna. Analisi delle fonti e problemi di metodo". In Brizzi and Varni Angelo, eds., *L'università in Italia fra età moderna e contemporanea. Aspetti e momenti, Bologna*, 1991, pp. 85-109.

[3] Mazzetti Serafino. *Repertorio di tutti i professori antichi, e moderni della famosa Università, e del celebre Istituto delle scienze di Bologna con in fine Alcune aggiunte e correzioni alle opere dell'Alidosi, del Cavazza, del Sarti, del Fantuzzi, e del Tiraboschi compilati da Serafino Mazzetti bolognese archivista arcivescovile*, 1848.
Brizzi Gian Paolo et al., *L'Università a Bologna*. Cassa di Risparmio in Bologna, 1988.

[4] Andrea Romano (curated by), *Gli statuti universitari: tradizione dei testi e valenze politiche.* Atti del Convegno internazionale di studi Messina-Milazzo, 2007.

[5] Gandolfi Giulia. "Imagines illustrium virorum: la collezione dei ritratti dell'Università e della Biblioteca Universitaria di Bologna", 2011.

Starting from these assumptions, this paper intends to describe how born digital primary sources could be used to reconstruct the online presence of the University of Bologna during the first twenty-five years of the World Wide Web[6]. The focus of this work is therefore primarily methodological: several different issues will be presented, starting with the fact that the University of Bologna website (hereinafter "Unibo.it")[7] has been excluded for thirteen years[8] from the Wayback Machine (Nanni, 2015), and possible solutions will be proposed and applied.

However, this research does not intend to be limited to what Blevins recently defined the "perpetual sunrise of methodology" (2015), which in his vision is the current main characteristic of digital history (a similar concept was also described in Gregory, 2014). In fact, the second aim of this work is to emphasize how these materials could give us new and distinct insights on the recent history of this academic institution, leading to interesting discoveries and becoming the starting point for several new historical analyses.

This paper is organized in three parts: in the first one, some of the fundamental concepts on web archives and the preservation of born digital sources will be introduced; moreover several important works related to this research will be described.

Then the reconstruction of the University of Bologna web's past will be presented. This was conducted using five different methodological approaches, namely: the use of Internet Archive materials, the employment of other web-archives platforms, the collection of documents preserved by the people who have managed the website in the last twenty years, the transcription of oral memories by the same people and finally the use of sources from other media (mainly national, local and digital newspapers).

[6] The case study here presented is part of the PhD research project that Federico Nanni has been conducting at CIS, The International Centre for the History Universities and Science of the University of Bologna (http://www.cis.unibo.it/), since October 2013. In his studies he is generally interested in understanding how the use of born digital documents will influence the historical method (with particular attention at the processes of retrieving, analyzing, selecting and visualizing sources).

[7] All the URLs mentioned in this research have been lastly checked on the 14th of April 2015.

[8] As it will be described later, between March and April 2015 (during the peer-review process) a collaboration with the Internet Archive was established. With the help of Christopher Butler and Giovanni Damiola, we discovered that the website has been excluded from the Wayback Machine since 2002.

Finally the future of this research will be described presenting a specific case study in which the historian's craft will be challenged by a completely different issue, namely the large amount of data available in the university digital library. As previously remarked, also in this case both the methodology that will be adopted and the research questions that will guide us will be presented and discussed.

## 1. Preserving the web of the past

In 1989 Tim Berners-Lee introduced its project at CERN, which later was identified as the "World Wide Web". In 1991 he created the first website, http://info.cern.ch/, and in the same year he publicly announced it in the Usenet newsgroup "alt.hypertext".

After a slow start[9], by the end of 1995 the web had more than 16 million users[10], which were, already at that time, the creators of a great amount of born digital traces. However the first project which focused on the preservation of this new kind of information started only at the end of 1996 under the leadership of Brewster Kahle. His already by then utopian purpose was of archiving the web in its entirety (Kahle, 1997).

The project he presented under the name of "Internet Archive" has become, during the last two decades, a fundamental archival work for the preservation of our digital past. Its crawlers started acquiring and preserving snapshots of webpages in November 1996, conducting an endless fight with the never-ending growth and continuous change of the web.

During the last twenty years several other platforms, often inspired by the ideas behind the Internet Archive but with a more specific national focus, have been developed, such as Pandora in Australia (1996), the UK Web Archive (2004), Netarkivet in Denmark (2005) and the Portuguese Web Archive (2007).

---

[9] Frana, Philip L. "Before the web there was Gopher". *IEEE Annals of the History of Computing*, 26.1, 2004, pp. 20-41.
[10] http://www.internetworldstats.com/emarketing.htm

Moreover, in 2003, the Internet International Preservation Consortium (IIPC), has been founded at the National Library of France[11]; during the last decade it has coordinated national and international efforts to preserve internet contents for the future. Today, with a General Assembly meeting every year since 2011 and organizations joining from 25 different countries, the IIPC has become the leading guide of these born-digital preservation projects.

## 1.1 The past of the Italian web sphere

Currently the National Libraries of Florence and Rome are not a part of the IIPC and no project with the specific purpose of preserving the Italian web-sphere exists. In 2006, thanks to the effort of the project "Crawler" (Bergamin, 2006; Tammaro, 2006) which was supported by the "Biblioteca Digitale Italiana" (Italian Digital Library), Italy cooperated with the European Archive Foundation (now called "Internet Memory Foundation") and conducted its first wide-spread crawling of the ".it" domain[12]. However, after this no other projects were conducted and the only part of its national web-sphere which has been constantly crawled and preserved are the PhD theses repositories of Italian universities, thanks to the activities of the Magazzini Digitali project.[13]

For all these reasons, as already described in Nanni (2014 and 2015), currently researchers interested in diachronically studying the Italian web sphere could rely only on the snapshots of websites preserved in the Internet Archive[14]. However, as Unibo.it has been excluded from the Wayback Machine[15] for more than a decade, this issue threatened to leave researchers with no primary sources related to the digital past of this academic institution.

---

[11] http://netpreserve.org/about-us
[12] This was conducted between May and June 2006, the snapshots are available here: http://collection.europarchive.org/bncf/
[13] http://www.depositolegale.it/
[14] Or, as it will be described later, in some specific cases on snapshots archived in other national web archives.
[15] The Wayback Machine is the Internet Archive platform that permits to display and browse the results of the crawl.

1.2 Decoding the message "This URL has been excluded from the Wayback Machine"

In order to understand the reasons of this removal, the first step of this research was to find, in the exclusion-policy of the Internet Archive, information related to the message "This URL has been excluded from the Wayback Machine", which appeared when searching "http://www.unibo.it".

In the FAQ section, the Internet Archive briefly describes the cases in which a website could be excluded by its platform[16]. The most common reason is when a website explicitly requests to not be crawled by adding "User-agent: ia_archiver Disallow: /" to its robots.txt file. As specified in the FAQs: "Alexa Internet, the company that crawls the web for the Internet Archive, does respect robots.txt instructions, and even does so retroactively. If a web site owner decides he / she prefers not to have a web crawler visiting his / her files and sets up robots.txt on the site, the Alexa crawlers will stop visiting those files and will make unavailable all files previously gathered from that site".
It is also explained that "Sometimes a website owner will contact us directly and ask us to stop crawling or archiving a site, and we endeavor to comply with these requests. When you come across a "blocked site error" message, that means that a site owner has made such a request and it has been honored. Currently there is no way to exclude only a portion of a site, or to exclude archiving a site for a particular time period only. When a URL has been excluded at direct owner request from being archived, that exclusion is retroactive and permanent".

When a website has not been archived due to robots.txt limitations a specific message is displayed. This is different to the one that appeared when someone searched the University of Bologna website, as you can see in Figure 1. Moreover, analyzing Google results related to "This URL has been excluded" message, it has been discovered that it was displayed in a few other cases, no one related to academic websites.

---

[16] https://archive.org/about/faqs.php#2

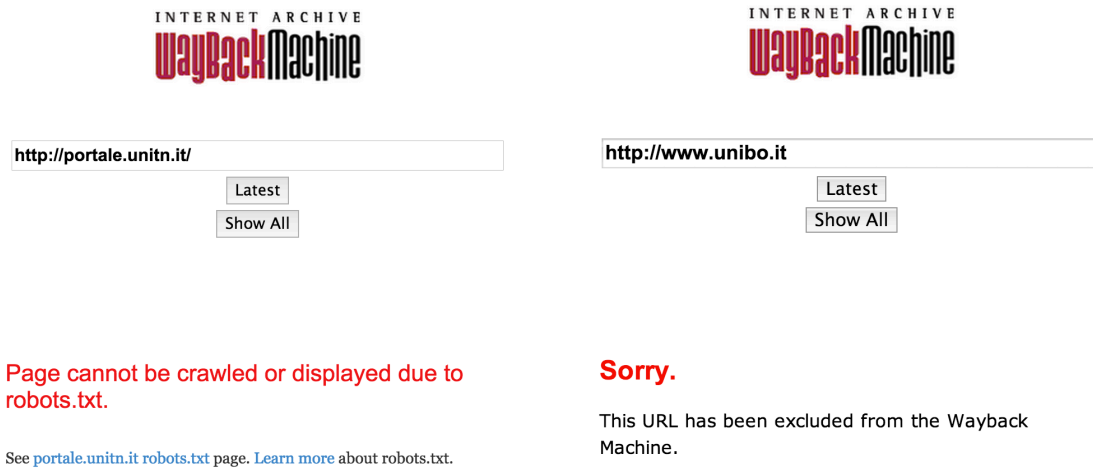

Figure 1. Here the two different messages are presented: the first one is related to the University of Trento website which has not been preserved on the Internet Archive between 2006 and 2010 due to robots.txt; the second one is related to the University of Bologna website.

Following the information offered by the Internet Archive in their FAQ section and assuming that "This URL has been excluded" message is what they define as "a "blocked site error" message", the only possible conclusion is that someone explicitly requested to remove the University of Bologna website (or more likely only a specific part of it) from the Archive.

For this reason we decided to consult CeSIA[17] (Area Sistemi Informativi e Applicazioni), the team which is currently managing the website, regarding this issue. However they did not submit any removal-request to the Internet Archive and they were not aware of anyone submitting it. To clarify this issue and discover whether the website of this institution has been howsoever preserved during the last decade, a collaboration between CeSIA and the Internet Archive has been recently established[18].

---

[17] http://www.unibo.it/it/ateneo/organizzazione/amministrazione-generale/1931/index.html
[18] After a series of unsuccessful attempts of contacting the Internet Archive team at info@archive.org (as it is suggested in the FAQ section) during 2014, a collaboration finally started at the end of March 2015. Thanks to the efforts of Mauro Amico (CeSIA), Raffaele Messuti

## 2. Studying the web of the past – related works

In 2010 Niels Brügger edited the first volume on a recent field of study which he defined "web history" (2010). Following that publication contributors on the topic (Ankerson, 2012; Brügger, 2012a; Roland and Bowden, 2012; Dougherty, 2013; Turkel et al., 2014) often highlighted the importance of two other papers that we could now identify as the starting points of the majority of these researches and speculations.

The first one, written by Roy Rosenzweig in 2003, is focused on how the work of historians will be more and more influenced in the future by both the scarcity and the abundance of materials (both digitized and born-digital). The second one, written by Brügger in 2009, remarks how the website has become an object of historical study itself, while underlining the importance of web-archival materials for historians.

With these works as the starting point, we could also identify two strongly connected areas of research that have generated publications in the field of web history during the last few years, one more focused on methodological approaches used to diachronically study the web (in order to deal in particular with the abundance of born digital sources) and the other more related to the use and reliability of sources preserved in the archives.

## 2.1 Computational web history

---

(AlmaDL), Christopher Butler (Internet Archive) and Giovanni Damiola (Internet Archive), the university main website became available on the Wayback Machine on the 13th of April 2015. However some URLs are still excluded, for example:
https://web.archive.org/web/http://estero.unibo.it .
As this result was achieved during the last part of the peer-review process of this paper, specific notes have been added to clarify its consequences on the future of this research. In particular, our first priority will be to discover the reasons of this exclusion-request, with attention to the possible correlation between it and the beginning of the "Portale d'Ateneo project" which, as it will be described later, started in that very same year (2002).
However the importance of this event is not limited to this specific case of study, but represents the risks of depending only on the efforts of an international organization such as the Internet Archive for the preservation of a national web sphere.

As one can notice by reading the submitted papers to the last Digital Humanities international conferences[19], computational textual approaches (which are often identified with the "distant reading" concept coined by Franco Moretti [2000 and 2013]) are one of the most important topics in the recent years of this heterogeneous area of research.

In particular the use of quantitative methods to analyze historical sources is experiencing a second wave[20], strongly sustained by natural language processing and text mining techniques, network analysis and machine learning approaches. These methods have been described and applied also in order to study the web of the past by Milligan (2012), Brügger (2012b), Hale et al. (2014), Samar et al. (2014) and in several interesting experiments described by Milligan on his blog[21]. Moreover, it is important to underline that, even though it was not a historical reconstruction, Foot et al. already in 2003 applied network analysis techniques to study the previous American electoral web sphere.

2.2 Reliability of Web Archives and Sources

A substantial number of articles focused on the reliability of web archives and web archival sources have been published in the last years. Howell in 2006 analyzed the Internet Archive and Murphy et al. (2007) established the Wayback Machine as a valid research tool. Brügger in 2008 underlined the need of what he called a "web-philology" to deal with the reconstruction of partially archived websites; more recently he also defined the materials preserved in web archives as "reborn digital materials" (Brügger, 2012c).

Dougherty et al. in 2010 summarized the state of the art of web archiving in relation to researchers and research needs; Ankerson (2012) underlined that web historians need to "consider broadcast historiography scholarship that grapples with questions of power, preservation, and the unique challenges of ephemeral media". Finally Ben-David and Huurdeman (2014) explored how to improve search tools in web archives and, with others (2014), employed a new approach to analyze hyperlinks in web archives in order to deal with the reconstruction of the unarchived web.

---

[19] http://dh2014.org/program/; http://dh2013.unl.edu/abstracts/; http://www.dh2012.uni-hamburg.de/conference/programme/abstracts/index.html
[20] After the first wave of quantitative history: Aydelotte, William Osgood et al. eds. "The dimensions of quantitative research in history". Princeton University Press, 1972.
[21] http://ianmilligan.ca/2015/02/03/using-modularity-to-explore-web-archives/

## 3. Setting up the research

This study, which is deeply influenced by the methodologies and the research questions highlighted in the works presented above, started from two important assumptions that are strongly related. The first one is that, as Brügger describes (2008 and 2011), web historians have to define and develop web-philological skills in order to reconstruct and understand the reliability of digital sources. Secondly, even if future historians will have to develop different theoretical and methodological approaches in order to analyze the structure or the content of a web-source of the past (as an example, studying the hyperlinked structure of a website or conducting a topic model representation of its textual materials), it will be more and more important to always keep in mind the intense bond between these two components (Brügger, 2002) and how they always influence each other.

### 3.1 Methodological approaches

As previously described, this research follows different approaches in order to retrieve digital sources related to the past of Unibo.it. First of all it has benefited from the work and expertise of the teams that currently manage the platform, in particular CeSIA, which also supervises the network infrastructure of the institution. This work also relies on the contributions and oral narratives of the technicians and the researchers who worked on the development of the platform in the past two decades.

Moreover, as the Internet Archive has not guaranteed a direct access to the website during the last decade, a key contribution came from materials retrieved from other national archives (in particular Netarkivet) that from time to time have preserved part of the platform[22].

Finally this work is also based on the information related to the university website that have been published on other media, such as local and national newspapers (in particular from the digital archives of *La Repubblica* and *Il Resto del Carlino*), university digital magazines (Unibo Magazine, Alma2000, AlmaNews), student forums (UniversiBo) and, in order to go further back in time, on Usenet discussions preserved on the platform Google Groups.

Currently the university still doesn't have an official Facebook page, but students' groups of the institution which are on this social network could be another useful source that will be explored more deeply in future. The same analysis will be conducted for university-related Twitter and Youtube accounts and from other materials published on social media platforms.

3.2 Analyzing structures and documents

Brügger (2009; 2012a) defined five different focuses in studying the web: the web element, the web page, the website, the web sphere and the web as a whole.

The presence of the University of Bologna online could be studied following each of these themes: as an example a study could deal with a methodological reconstruction of its homepage (a specific webpage), or as in Hale et al. (2014), in the analysis of its role in the hyperlinked structure between Italian universities (a web sphere).

Therefore, even if the focus of this research will be specifically related to the evolution of the principal components of its structure (i.e. the role of its homepage in relation with the department pages), other focuses as the ones mentioned in the previous paragraph will be considered, in order to achieve a better understanding of the digital role of the platform.

## 4. The history of Unibo.it: 2015-2002

---

[22] As described in Nanni (2015) this is due to how the archive-crawler is set (how many hyperlinks it will follow starting from a specific point).

The majority of the information published on the website between the early 2000s and 2015 are still available online (for example all the courses programs, the descriptions of research projects and the contracts and grants published by each School[23]). However in order to retrieve them researchers rely on the results of a general "string-matching" search tool[24].

Moreover, as previously mentioned, the Internet Archive has not offered any snapshot of the homepage of Unibo.it for the last thirteen years. Therefore, during this research it has not been possible to monitor how the layout changed during time or analyze the impact of Italian school reforms on the structure of the platform.

4.1 The website as it is structured today

The website of the University of Bologna is currently offered in two different versions: an Italian one (which is available at the URL: http://www.unibo.it/it), and an English one. As the English version offers the translation of only a part of the website, the focus of this research will be mainly on the Italian one. However in the future it could be also interesting to conduct a comparative analysis of university websites in their own language and in the translated versions, to better understand, as an example, which parts are translated and which not[25].

Moreover, as the university is divided in five campuses, the website is consequently divided into five subsections (for example, http://www.unibo.it/it/campus-forli); however, this study will follow only the evolution of the main one.

This platform is currently managed by two different offices: as mentioned above "CeSIA - Settore Tecnologie web" takes care of the structure (called "Sistema

---

[23] Since the academic year 2012/13 the 23 faculties have been reorganized in 11 schools.

[24] http://search.unibo.it/UniboWeb/UniboSearch/Default.aspx

[25] This could highlight for example specific practices of "academic web-marketing".

Portale di Ateneo", as it will be described later), and "AAGG — Ufficio Portale Internet e Intranet di Ateneo"[26] manages the contents.

Using the XML Site-Map Generator[27] the several subsections of the website have been mapped. These are:

- /ateneo/ : general information on the university, its organization, its administration, its structure (schools, departments, etc).

- /bacheca/ : news and other general information which also appears in other subsections (i.e. news related to education, to research, etc).

- /didattica/ : educational information (bachelor, master degree and PhD programs, summer and winter schools, etc.).

- /enti-imprese/ : information for public and private companies.

- /futuri-studenti/ : information for future students.

- /internazionale/ : information on international collaborations with other universities (Erasmus, Marco Polo and Overseas programs)

- /laureati/ : information for graduated students.

- /personale/ : information for university employers.

- /ricerca/ : research information (patents, spin-offs, projects for the 2015 EXPO)

- /servizi-e-opportunita/ : different services for the user (i.e. how to connect to the university wi-fi, how to have access the personal e-mail, information related to scholarships, etc).

- /studenti-internazionali/ : information for enrolled international students.

- /studenti-iscritti/: information for enrolled students.

In addition to this, the website has several sub-domains (more than 90 were found by an exploratory study using Find Subdomains - Pentest Tools[28]), which offer different services (a platform called AlmaOrienta which helps high-school students in choosing between the different bachelor degrees), digital publications (for instance the Unibo Magazine), beta testing platforms

---

(beta.unibo.it) and old abandoned department sub-domains (http://www2.stat.unibo.it/; http://www2.classics.unibo.it/).

It is important to remark that in the early steps of this research we discovered that, even if the university of Bologna homepage and all of its subsections were not accessible through the Wayback Machine, its sub-domains were available on the Internet Archive and have been constantly preserved during the last twenty years (i.e. Unibo Magazine:
 https://web.archive.org/web/*/http://magazine.unibo.it).

4.2 The changing moment

Luca Garlaschelli was the Chief of Information/Innovation Office (CIO) at the University of Bologna between 2002 and 2012. Under his supervision the "Sistema Portale di Ateneo" was created[29]. This is a general interface to a hierarchical organization of all the digital resources of the university that are available online[30]. As it will be presented in the next pages, this has led to a revolutionary transformation of the digital presence of the institution[31].

Among several important improvements, this project required that all departments and degree web pages change their structure and adopt a common layout and organization. As examples, the department of Classic and Medieval Philology and the department of Computer Science had to change their URL addresses to standardize ones ("abbreviation of the name of the department" + "unibo.it"), the first one from: http://www.classics.unibo.it/ became

---

[29] As described here: http://www.slideshare.net/lucagarlaschelli/private-cloud-computing-in-organizzazioni-complesse
[30] As underlined by Garlaschelli in an interview in 2004:
http://www.osservatoriosullacomunicazione.com/mezzi/internet/prontoweb/interviste/garlaschelli.php.
And also presented in the "Annuario degli anni accademici 2003-2004 e 2004-2005", pp. 747-750: http://www2.unibo.it/Annuari/Annu030405/Annuario0304-0405.pdf
[31] For three consecutive years the website received the "Osc@r del web" prize as the best Italian p.a. digital platform: http://www.magazine.unibo.it/archivio/2007/oscar_del_2007. In 2007 Luigi Nicolais, the Italian Minister of Public Administration, was also present to confer this honour.

http://www.ficlit.unibo.it/ and the second one from http://www.cs.unibo.it/ became http://www.informatica.unibo.it/.

This transition started in 2004 and even if some departments decided to keep the older version of their sub-domain online by adding a "2" after the "www"[32] (as an example, the previously mentioned http://www2.classics.unibo.it/), for the majority of them the old versions of their sub-domain is no longer available on the live web (as an example the department of History, whose URL was: http://www.dds.unibo.it/).

As previously described, it is still possible to retrieve a sub-domain of the University of Bologna website on the Internet Archive, but only if the original URL is known, which is not a trivial issue (as described in Nanni, 2015).
As an example, the Department of Philosophy and Communication Studies was divided in two different departments until 2012 and the department of Philosophy used http://www.filosofia.unibo.it as a URL even before the transition to the "Sistema Portale d'Ateneo". However, in the Nineties, this department used for a couple of years another URL: http://www.sofia.philo.unibo.it which, without the memories of the people that managed the sub-section at that time, would have been be very difficult to discover.

Summarizing, we could identify a specific turning point in the history of the University of Bologna website. With the "Sistema Portale d'Ateneo" project and in particular with the standardization of departments pages which started in 2004[33], the website has been completely re-organized and the majority of the previous contents on these pages have been deleted from the live web. However if they maintained the same URL or if the previous URL is known[34], these materials and their structure can be retrieved from the Internet Archive.

---

[32]  This could be due to a personal choice of the person who was managing each department page at that time and not to a decision of the CIO. On the department of Classic and Medieval Philology homepage it is explicitly written that "the pages will continue to be available, but will be no longer updated".
[33] http://www.magazine.unibo.it/archivio/2004/dipartimenti

4.3 Reconstructing the changes of the interface

Mauro Amico, head of the web-technologies department at CeSIA, offered a collection of seven .png images (see Appendix) that capture the most important istances in the evolution of the organization of the homepage before the current layout[35].

Looking at the ones after 2002 we can notice that, even if a few graphical adjustments were conducted (the Unibo-Magazine was introduced on the left in 2004; the search tool was repositioned in the center in 2006, etc.), the structure remained more or less the same until the July 2013, when the current interface was presented.

The present organization of the "Sistema Portale d'Ateneo" is the first one completely created by CeSIA without the supervision of Luca Garlaschelli and, along with a new graphical interface, its main characteristic is that it offers for the first time the possibility of surfing the website as a specific user (as a prospective student, a student, a private company, etc.) and it proposes different contents accordingly.

Even if these .png images give us a first idea of the different interfaces, in order to become able of exploring again the old versions of the website, other services have to be employed.

As a start, the Internet Memory offers the results of the 2006 national ".it" crawl online, but only a single snapshot of Unibo.it homepage is available (archived on the 8th of May 2006[36]) and is not even well preserved. However, as previously remarked, other national web archives have captured from time to time parts of the Italian web sphere and, among them, Netarkivet preserved Unibo.it several times between 2006 and 2012.

---

[34] A web page archived in 2002 could help us identify the URL of each department in that year: https://web.archive.org/web/20020224030346/http://alma2000.unibo.it/facolta/dipE.asp
[35] They cover the periods: 01/1996-01/1998; 01/1998 – 09/1998; 09/1998-07/1999; 2002-2003; 2004-2206; 2006-2009; 2009-2013. As it will be remarked later, there are no images related at the period: 1999-2002.

Moreover the Internet Archive completely preserved the English version of the website between 2004 and 2014[37] and also the Chinese version of the platform between 2006 and 2013[38]. In 2014 the English version, as it has become a subsection of the main website (its URL changed from http://www.eng.unibo.it to http://www.unibo.it/en/homepage), has been excluded by the Wayback Machine too.

Both the snapshots preserved in Netarkivet and the English version of the website in the Internet Archive could finally give us the possibility of surfing Unibo.it in order to examine more in details its structure and, as it will be described in the next pages, to retrieve interesting primary sources[39].

4.4 News from its recent past

The use of primary sources from the digital archive of the newspaper *La Repubblica*[40] has been a consistent help in this study. Through these articles it was for example discovered that in 2003 the university introduced, on its website, the digital edition of the student-guide of the city of Bologna, as also described in a news on the Unibo Magazine[41]. This guide was written by Umberto Eco, Carlo Lucarelli and other famous professors and writers.

---

[36]
http://collection.europarchive.org/bncf/20060508021404/http://www.unibo.it/Portale/default.htm
[37]
https://web.archive.org/web/20041023113357/http://www.eng.unibo.it/PortaleEn/default.htm
[38]
http://web.archive.org/web/20070222051105/http://www.eng.unibo.it/PortaleEn/ChinaAssociationCollege_cn.htm
[39] It is important to remark that, before solving the exclusion-issue in April 2015, this was the only way of browsing the old versions of the website.
[40] http://ricerca.repubblica.it/repubblica/archivio/repubblica/2003/09/18/universita-insegna-la-dolce-vita-da.html?ref=search
[41] http://www.magazine.unibo.it/archivio/2003/09/22/guida-di-bologna

Even if the digital version is not available online anymore and its URL has not been preserved in any of the platforms previously mentioned, a link to a "guide to the guide" is still available online. This document shows a group of useful digital resources for new students, i.e. the platform "Flash Giovani", created with the support of the municipality of Bologna and focused on the cultural activities in the city and the website "Studenti.it" which has become in the last fifteen years one the most important Italian online communities for high school and university students[42].

As described in Unibo Magazine[43], since 2004 each professor has had a personal page, in which she/he publishes the course programs (with links to several other digital materials), her/his research interests, and is able to update the publications list. This specific kind of documents could be really useful in the future to conduct a large analysis on which have been, year after year, the most important topics in educational programs. This could also enable us to evaluate the influence of specific scientific discoveries (how many courses mentioned "Higgs boson" before 2013? And after?). It is important to underline that professors main pages were also excluded from the Wayback Machine and just a few of them have been preserved by other national archives.

Digital sources related to the recent years of the university allowed us to discover how in 2005 the future "Prorettore per la ricerca" Dario Braga underlined the importance of starting to teach courses in English (and also Chinese and Arabic) among his "proposals for the future"[44] or how five years later, during his administration, he discussed[45] in a Google Group newsletter the impact of the "Gelmini" school reform with a group of professors named "Docenti preoccupati" ("worried professors")[46].

---

[42] Currently it is one of the 200 most visited websites in Italy:
http://www.alexa.com/siteinfo/studenti.it#trafficstats
[43] http://www.magazine.unibo.it/archivio/2004/pagina-personale-docente/
[44] http://ricerca.repubblica.it/repubblica/archivio/repubblica/2005/04/13/proposte-per-ateneo-del-futuro.html?ref=search
[45] https://groups.google.com/forum/#!topic/docentipreoccupati/WqOJqQzkPmU
[46] Another interesting source in order to study the experience of Dario Braga as Prorettore and its run for the future Rettore of the university, will be his personal blog:
http://www.dariobraga.com/blog

Moreover, these materials gave us information on the activities of the "Centro Studi La Permanenza del Classico" directed by the current Rettore, Ivano Dionigi[47], showed how the Unibo Magazine presented itself online in 2003 (with an interview[48] of the then Rettore, Pier Ugo Calzolari, which focused on the scarcity of funding for higher education and research in Italy) and revealed which were the problems on the usability of the website for people with disabilities[49] .

Among all these different resources one piece of news specifically captured our attention: in May 2007 a group of activists decided to create a copy of the Unibo.it interface. They were protesting against the European Credit Transfer and Accumulation System (ECTS) for the evaluation of the amount of hours of study. At the URL http://www.unibologna.eu/[50] an identical version of the homepage was available, with the description of the reasons of the protest.

In a couple of weeks the website attracted a high number of visitors and most of all the attention of the university[51], which blocked the access to the platform from all its computers[52].

This source is not only important in our study as it documents a different and innovative way of conducting a protest against this institution, but as the fake-website has been preserved by the Internet Archive it also offers a completely preserved version of the layout of the Unibo.it homepage.

## 5. The history of www.unibo.it: 2002- the early Nineties

In order to reconstruct the first ten years of history of this platform neither material on the live web nor documents in other national web archives are available. For this reason, the second part of this study will firstly employ information from local and national newspapers, which often described new services offered online by the university.

---

[47] http://ricerca.repubblica.it/repubblica/archivio/repubblica/2006/10/25/umanisti-scienziati-insieme-alla-stessa-lezione.html?ref=search
[48] http://www.magazine.unibo.it/archivio/2003/11/11/intervista-al-rettore
[49] http://ricerca.repubblica.it/repubblica/archivio/repubblica/2006/12/10/almawelcome-non-vedenti.html?ref=search
[50] Later moved to: http://www.unibologna.eu/www.unibo.it/Portale/default.html
[51] http://www.magazine.unibo.it/archivio/2007/attacco_al_portale
[52] http://ricerca.repubblica.it/repubblica/archivio/repubblica/2007/06/13/clonato-il-sito-dell-ateneo-per-protesta.html?ref=search

Moreover, in the attempt to discover when this platform was created, by whom and for which initial purpose, two different web-server lists will be analyzed.

5.1 Different ways of going back in time

In order to study the structure of the website before the "Sistema Portale d'Ateneo" several different sources will be employed, which will in turn help us in understanding how the website looked like, how it was used and how relevant it was in the academic digital ecosystem.

In particular, as we described earlier, the archive of the newspaper *La Repubblica* offers several news on how the platform changed during the Nineties. For example, we discovered that the institution offered a free email account to all students since 2002[53] and it was the first Italian university which gave the possibility of paying fees online (2000) [54]; moreover since 1999 some departments also guaranteed the possibility of enrolling online to courses and exams[55].

Another interesting piece of news retrieved from the digital archive of *La Repubblica* is from October 2001, a few months before the project "Portale d'Ateneo" started. In those days the University of Bologna website won the "WWW" prize from the Italian economic newspaper "Il Sole 24 Ore" for the best website in the category "School, university and research". At the ceremony Salvatore Mirabella, a technician who managed the website during the Nineties, was also present[56]. However, as the snapshot of that homepage was not in the collection offered by CeSIA, during this research was not possible to retrieve any image of the layout of the website in those years. In the near future Mirabella will be also consulted in order to improve our knowledge on the history of the platform.

---

[53] http://ricerca.repubblica.it/repubblica/archivio/repubblica/2002/10/09/mail-gratuita-per-gli-studenti.html?ref=search
[54] http://ricerca.repubblica.it/repubblica/archivio/repubblica/2000/07/17/tasse-online-per-universita.html?ref=search
[55] http://ricerca.repubblica.it/repubblica/archivio/repubblica/1999/09/13/finalmente-offerta-portata-di-mouse.html?ref=search
[56] He was the head of "Urp – Servizio Web" , as described here:
http://www2.unibo.it/Annuari/Annu9901/Indice/parte2/parte2sez1/parte2sez1.html

However, as we can notice by looking at the images offered by CeSIA or analyzing a few examples that are still available on the live web (i.e. http://www2.unibo.it/annuari/ or http://www2.unibo.it/avl/english/default2.htm)[57], before the "Sistema Portale d'Ateneo" project the homepage of Unibo.it was mainly an information page, with only a few general links.

At the same time, consulting "The list and map of the Italian WWW servers"[58] created by Cilea and available since 1997 on the Internet Archive[59], we can notice how several departments, faculties and research groups were already online and, as opposed to the relatively passive homepage, very active in the Nineties. We retrieved, as an example, all the information on courses in History since 1998[60], the organization of the university astronomical observatory[61] and of the faculty of Engineering[62] since 1997, description on the inter-faculty library since October 1996[63] (the entire system was created in 1993[64]), the digitization of the students guide books carried out by the faculty of economics in 1994[65].

These different pages were continuously updated with new information by technicians, researchers, professors and, from time to time, also with contribution of the students[66].  For these reasons, they all evolved differently during the Nineties and they are now interesting artifacts on how the departments of this university approached the World Wide Web.

---

[57] It is important to notice that the page http://www2.unibo.it is not available on the live web anymore and it was excluded from the Wayback Machine.
[58] This is a useful starting point for every researcher who is interested in the past of the Italian web sphere.
[59] http://web.archive.org/web/19971025045601/http://www.cilea.it/WWW-map/
[60] https://web.archive.org/web/19981206110539/http://www.dds.unibo.it/
[61] http://web.archive.org/web/19970114105744/http://www.bo.astro.it/
[62] https://web.archive.org/web/19970422153341/http://www.ing.unibo.it/
[63] https://web.archive.org/web/19961031164155/http://liber.cib.unibo.it/
[64] https://web.archive.org/web/20010424195903/http://www.cilea.it/collabora/GARR-NIR/nir-it-2/atti/cib.html
[65] https://web.archive.org/web/20010424195734/http://www.cilea.it/collabora/GARR-NIR/nir-it-2/atti/didbo.html
[66] https://web.archive.org/web/19980117121355/http://caristudenti.cs.unibo.it/index.shtml

However, there are many other ways of looking at the second half of the Nineties history of Unibo.it. One approach could be to study the information related to AlmaNET, the university Internet connection, built in 1988 and highly improved in 1996[67] with the collaboration of Telecom Italia and under the supervision of CeSIA (established in 1994). Another way could be to conduct a specific research on the adoption, at the level of the university, of the service AlmaLaurea in May 1998[68], which aimed at improving the relationship between the academia and the job-market, or to discover how young students used Usenet to talk about the different degrees offered (for example on the distinction between "Computer Engineering" and "Computer Science"[69]).

The Internet Archive has also preserved Unibo.it's old online magazine, AlmaNews, which offered several short videos of important events, such as the ceremony[70] for the first degrees in Business Administration and Political Sciences in 1997.

It is evident that studying the digital past of this institution and using these resources to improve our knowledge on its recent history is a complex challenge that relies on both new methodological approaches and specific research questions. Even if the few examples presented here have the only aim of highlighting different ways of looking at its digital past, the reconstruction process itself could guide us in the research.

As an example, it is known (Chiara, 1998) that the city of Bologna and its citizens have a strong bond with innovation in computing technologies (the municipality created one of the first civic-networks in the world in 1995, giving to all citizens

---

free access to the Internet the very next year [Chiara, ibidem]). This relation emerged also in an article of 1996 founded in *La Repubblica* digital archive.

In November of that year, for the first time in Italy, a digital discussion was censored and an entire mailing list named "Lisa" was completely closed. This happened on the Unibo server[71]: CeSIA informed the professor of computer science Dario Maio of the presence of violent debates on the platform and the department of Computer Science decided to drastically intervene.

As the article reported, these digital conflicts were probably related to the internal discussion of an Italian association denominated "La città invisibile"[72] (The invisible city). This association, comprising early Internet activists, was interested in sharing the importance of digital cultures and rights. Among them there were also academics, for example Lucio Picci (currently professor of Political Economy[73]), who was at that time a young researcher at the University of Bologna[74].

## 3.4 At the beginning of the digital era

Tracing the first online presence of an entity such as the University of Bologna is not an easy task. First of all it is not possible to use a tool such as Domaintools.com[75] as it has records only until 1995. In Italy a list of .it server was initially maintained by the research center Cnuce and currently is available on the platform Registro.it[76]. However all early Italian websites (created before 1996) have a common creation date: 29-01-1996 (Figure 2).

| Dominio | | Dominio | |
|---|---|---|---|
| **Dominio:** | unibo.it | **Dominio:** | crs4.it |
| **Stato:** | ok | **Stato:** | ok |
| **Data Creazione:** | 29-gen-1996 0.00.00 CET | **Data Creazione:** | 29-gen-1996 0.00.00 CET |
| **Data Scadenza:** | 2-mag-2015 CET | **Data Scadenza:** | 29-gen-2016 CET |
| **Data Aggiornamento:** | 18-mag-2014 0.53.11 CET | **Data Aggiornamento:** | 14-feb-2015 0.52.48 CET |

---

[71] http://ricerca.repubblica.it/repubblica/archivio/repubblica/1996/11/24/troppe-parolacce-censurata-lisa.html?ref=search

[72] http://www.citinv.it/intro.html

[73] http://www2.dse.unibo.it/picci/

[74] http://www2.dse.unibo.it/picci/sfera/storia.html

[75] http://www.domaintools.com/

[76] http://www.nic.it/

Figure 2. Unibo.it and Crs4.it (the first Italian website, created in 1991) have the same starting date.

The research group "GARR-Network Information Retrieval" organized a series of annual meetings in the early Nineties[77], dedicated to the diffusion of the World Wide Web in Italy at the university level.

Consulting the proceedings of 1994 we learned[78] that Unibo.it was already active at least in the August of the year before (when the digital library environment project, which is the subject of the paper, started); however, unfortunately we cannot consult the proceedings of the first meeting (1993), as they were shared on Gopher (gopher://itocsivm.csi.it/11/INTRCONV) and for the moment are not at our disposal. Another resource that will be explored extensively in the future will be the bibliography dedicated to "Internet in Italy" edited by Riccardo Ridi on its website in 1997[79].

During the first years of the World Wide Web Tim Berners-Lee curated a list of web-servers on the CERN website; the last update available is from late 1992[80]. Both Unibo.it and Crs4.it[81] are not mentioned in this list, but there is a link to another Italian research institution, the Physics Institute in Trieste.

Later, on the NCSA website, a specific section called "What's New!" published each month (from June 1993 to January 1996) a list of the new servers on the web. No search tool is available, but consulting each page we found some interesting information about specific sub-sections of Unibo.it, for example the "Bologna Astrophysics Preprints" has offered online since November 1994 all the scientific publications of the Bologna Astronomical Observatory (OAB), the Astronomy Department of Bologna University (DDA), the Radioastronomy Institute of CNR (IRA) and the TESRE Institute of CNR (ITE).

---

[77] http://web.archive.org/web/19971025045540/http://www.cilea.it/GARR-NIR/
[78] http://web.archive.org/web/20010424195903/http://www.cilea.it/collabora/GARR-NIR/nir-it-2/atti/cib.html
[79] http://www.riccardoridi.it/esb/biblint/04.htm
[80] http://www.w3.org/History/19921103-hypertext/hypertext/DataSources/WWW/Servers.html
[81] In the future it could be interesting to discover the reasons of this specific exlcusion.

However, for what concerns specifically the creation date of the website, in December 1993 a link to a map of all Italian web-server was published, but this link is not available anymore (it redirects to the 1997 version of the Cilea Map).

Summarizing, for the moment we know that, as the website of the research center "Centro di ricerca, sviluppo e studi superiori in Sardegna" (CRS4 – www.crs4.it) is considered the oldest one in Italy (it was created in 1991[82]) and that Unibo.it was already available in the second half of 1993, the website had to be created between these two dates, more precisely after the end of 1992 according to CERN web-server list.

Future researches on this matter will involve the consultation of Salvatore Mirabella and Renzo Davoli[83] in order to discover who created the html structure of the website and for which initial purposes, the retrieval of the proceedings of the first "GARR-Network Information Retrieval" meeting and the analysis of university archived records[84] which could be related to the initiation of the web project (such as bills from the purchase of computers, to be used as web servers[85])[86].

## 4. Dealing with the abundance of sources

As mentioned in the introduction, the only part of the Italian web-sphere which has been constantly crawled and preserved during the years are the repositories of PhD theses published by Italian universities[87]. Currently it is not possible to have access to this "National PhD Theses Database", however several Italian universities guarantee an open access to their collections.

---

[82] Pinna A., "Soru: un incontro con Rubbia, così nacque il web in Sardegna", Il Corriere della Sera, 28/12/1999, p.24
[83] In the early Nineties he was the person in charge of the University TCP/IP network.
[84] Another source that will be more explored in the future are the university yearbooks of the period 1990 – 1993, in which the creation of the website could have been mentioned.
[85] It is important to remember that the website was on CeSIA servers during the second part of the Nineties, but CeSIA itself was established only in 1994.
[86] On the 23th of April 2015, during the last days of the peer-review process, we had a meeting with Renzo Davoli (professor of Computer Science at the University of Bologna). Thanks to his help and the collaboration of his colleague, Ozalp Babaoglu, we discovered that the web domains "cs.unibo.it" (Computer Science) and "dm.unibo.it" (Department of Mathematics) were registered in July 1993. The future of this research, in particular for what concerns the first years of Internet in Bologna, will rely on a solid collaboration with the scholars here mentioned.
[87] http://www.depositolegale.it/deposito-legale-digitale-delle-tesi-di-dottorato/

Specifically on Alma Digital Library (AlmaDL - which was created between 2001 and 2002[88]), the University of Bologna has offered during the years several different resources (educational materials, historical documents, etc).

However, as will be explained in the next section, this enormous corpus of documents (which since 2008 has been composed by PhD theses and later by Bachelor and Master theses[89]) presents completely different issues for a historian interested in employing digital resources to reconstruct the University of Bologna's recent past.

## 4.1 The PhD-Corpus

If we focus our attention on the Phd Corpus of AlmaDL[90], we will find over 4500 theses at our disposal. During the years, they have been preserved both by Magazzini Digitali and by the Internet Archive. Therefore, following Rosenzweig's (2003) definition, the main issue that historians interested in employing these materials will encounter here is the abundance of digital sources.

In fact for the moment the only access to the archive is guaranteed by a search tool which allows to perform a string-matching query on the metadata (i.e. the title, the author, the abstract, etc). Moreover the theses are divided following the Italian academic eco-system, which is organized in 14 broader research-areas (Mathematics and Computer Science; Physics; History, Philosophy, Pedagogy and Psychology, to name a few) and in more than 200 specific disciplines (such as Demography, Maritime Law and Plant pathology). However this organization is not helpful if someone is interested in understanding which are the most recurrent research-topics for a general discipline (such as History), how these

---

topics changed year after year or if these documents could give us an insight of interdisciplinary collaborations between disciplines that highlight new research topics and interrelations between departments.

4.2 Methodological approach

As already remarked by Ramage et al. (2011) corpora of Phd theses could give researchers "an excellent basis for the study of the history academia because they reflect the entire academic output of universities, as seen through their graduating students, and do not reflect the coverage biases toward scientific or engineering publications found in most databases of academic publications, such as ISI (with a biomedicine) and CiteSeer (with computer science)".

Following their approach (described also in Chuang et al., 2012), which was employed on the Stanford doctoral dissertation corpus, a similar study will be conducted on the Unibo corpus[91]. This research will face new specific issues, such as a multi-lingual analysis of the corpus (the abstract and the theses are mainly in English and Italian) and the absence of a specific discipline-label (the sub-sections are too many and several of them recur only once)[92]. In order to find similarities between these documents and extract the most recurrent research topics different methods will be employed (a cosine similarity comparison using the tf-idf representation of each document, a topic model distribution, etc).

Finally in order to detect interdisciplinary works, as common methods (such as citation analysis)[93] could not be employed because the references of the theses are not available, both the potentialities of a topic model representation of the corpus and of other non-supervised classification methods (combined for example with the detection of outliers) will be tested.

---

[91] This specific project will be conducted during 2015 in collaboration with the Data and Web Science Group of the University of Mannheim (under the supervision of Simone Paolo Ponzetto) and it will represent the second part of Federico Nanni's PhD research project. We decided to mention it in the last part of this paper in order to stress how the same case study could challenge the historical method under a completely different perspective.

[92] Ramage et al. use the department affiliation of the supervisor as a discipline label. This solution will be experimented, however recently some departments at the University of Bologna have been merged and therefore this could not be an absolute indicator.

[93] For example: Leydesdorff, Loet, and Ismael Rafols. "A global map of science based on the ISI subject categories." *Journal of the American Society for Information Science and Technology,* 60.2, 2009, pp. 348-362.

## 5. Conclusions

The aim of this paper was to highlight both the issues and the potentialities of using born digital documents to study the recent past of the University of Bologna. The main focus of this research was to describe the methodological approach employed to conduct a historical reconstruction of its website (which has been excluded for the last thirteen years from the Wayback Machine).
In doing so we underlined how its history is divided into two parts (before and after the advent of the "Sistema Portale d'Ateneo" platform) and how different sources (materials from other archives, document preserved by CeSIA, articles on local, national and digital newspaper) have been useful to improve our knowledge on the evolution of this platform (in particular on the role of department pages). This work also remarked how born-digital sources can offer new insight on common research topic related to the history of this university and its relation with the students' community and the city itself (we summarized the results of this study in a timeline – Appendix 2).

The final parts of this work were focused on two specific research issues that remind of the "scarcity and abundance" of digital materials described by Rosenzweig (2003). On one hand the difficulties of going beyond November 1996 (the month when the Internet Archive started preserving the web) in order to discover the first years of this university online were presented. On the other hand, a different issue was introduced: the University of Bologna since 2002 has been collecting resources on its digital library platform, and since 2008 a corpus of PhD theses has been made available by the institution.

These two different issues highlight the necessity of a different methodological approach for the new generation of historians. This modification of the historian's craft is not only related to the use of computational tools and methods, but following what Brügger defined as web-philology (2008) and

Dougherty recently evoked as web-archeology[94], it relies on a new way of conceiving the retrieval, analysis and employment of primary sources which will sustain traditional historical research questions and will lead to a infinite number of new ones.

## 6. Acknowledgements

---

[94] http://www.netlab.dk/projects/p8-virtual-digs-excavating-preserving-and-archiving-the-web/

**Appendix 1**

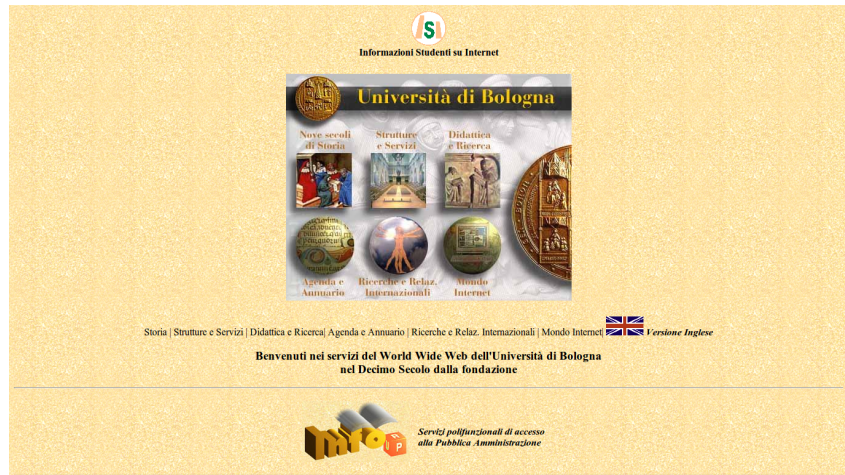

01/1996 – 01/1998

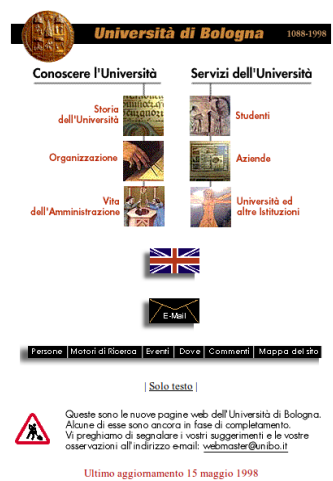

01/1998 – 09/1998

09/1998 – 07/1999

2002 - 2003

2004 – 2006

2006 - 2009

2009 – 07/2013

07/2013 - …

(Snapshot taken on the 17th of March 2015)

## Appendix 2

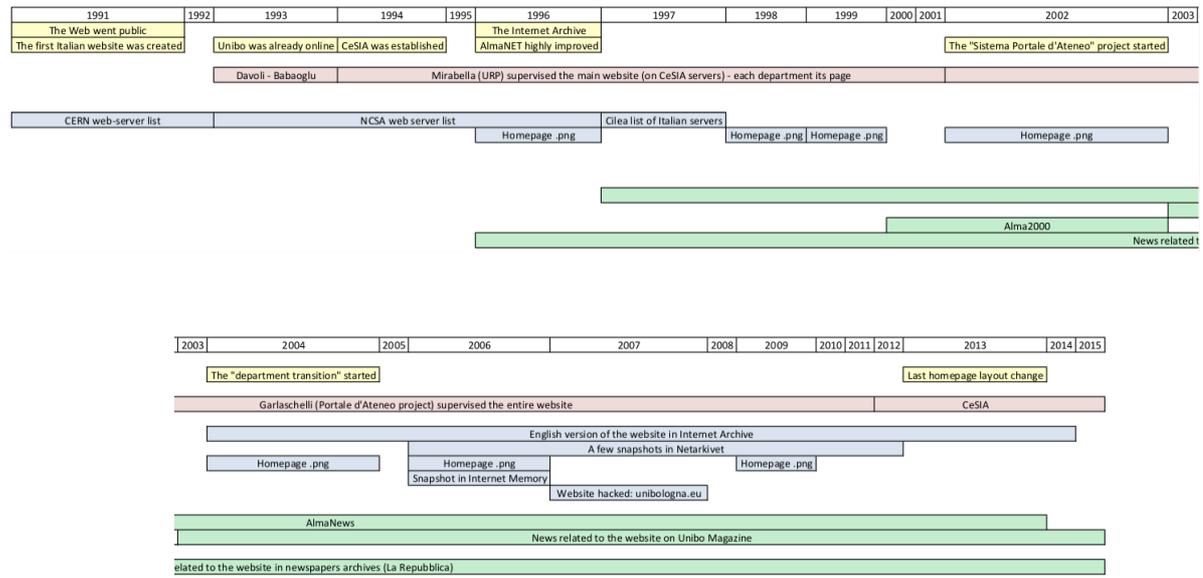

In this figure we present a few important historical events related to the Unibo.it web presence (yellow), who managed the website during the last two decades (red) and two different kinds of sources available, direct ones (i.e. snapshots or link to the website in server lists - blue) and indirect ones (i.e. news – green).